\documentclass{aa}
\usepackage[varg]{txfonts}

\usepackage{txfonts}
\usepackage{graphicx}
\usepackage{inputenc}
\usepackage{natbib}
\usepackage{color}

\newcommand{\degree}{\ensuremath{^\circ}}
\begin{document}

\title{The Luminosity Function of Low Mass X-ray Binaries in the Globular Cluster System of NGC 1399}
\author{G. D'Ago\inst{1,2,3}, M. Paolillo\inst{1,4,5}, G. Fabbiano\inst{6}, T.H. Puzia\inst{7},  T. J.Maccarone\inst{8,9}, A. Kundu\inst{10,11}, P. Goudfrooij\inst{12}, S. E. Zepf\inst{13}}
\institute{
Physics Dept., University of Napoli Federico II, via Cinthia 9, 80126, Italy; \email{paolillo@na.infn.it}
\and
Physics Dept. "E.R. Caianiello", University of Salerno, via Ponte Don Melillo, 84084, Fisciano (SA), Italy
\and 
INFN, Gruppo collegato di Salerno, Sezione di Napoli, Italy
\and
INFN - Sezione di Napoli, via Cinthia 9, 80126, Napoli, Italy
\and
ASI Science Data Center, Via del Politecnico snc, 00133 Rome, Italy
\and
Harvard-Smithsonian Center for Astrophysics, 60 Garden St., Cambridge, MA 02138, USA
\and
Institute of Astrophysics, Pontificia Universidad Cat\'{o}lica de Chile, Avenida Vicu\~{n}a Mackenna 4860, Macul, Santiago, Chile
\and
Department of Physics, Texas Tech University, Box 41051, Lubbock TX, 79409-1051
\and
School of Physics and Astronomy, University of Southampton, Southampton SO17 1BJ, UK
\and
TIFR, Homi Bhabha Road, Mumbai 400005, India
\and
Eureka Scientific Inc., 2452 Delmer St, Suite 100, Oakland, CA 94602, USA
\and
Space Telescope Science Institute, Baltimore, MD 21218, USA
\and
Department of Physics and Astronomy, Michigan State University, East Lansing, MI 48824, USA
}

   \date{Accepted on May 1 2014 for publication on A\&A}

\authorrunning{G. D'Ago et al.}
\titlerunning{Faint LMXBs in NGC 1399}

\abstract{}{We present a study of the faint-end of the X-ray Luminosity Function (XLF) of Low Mass X-ray binaries (LMXBs) in the Globular Cluster (GC) system of the cD galaxy NGC 1399.}{We performed a stacking experiment on 618 X-ray undetected GCs, in order to verify the presence of faint LMXBs and to constrain the faint-end slope of the GC-LMXBs XLF below the individual detection threshold of $8\times10^{37}$ erg s$^{-1}$ in the $0.5-8$ keV band.}
{We obtain a significant X-ray detection for the whole GC sample, as well as for the red and blue GC subpopulations, corresponding to an average luminosity per GC $<L_{X}>_{GC}$ of $(3.6\pm1.0)\times10^{36}\ erg\ s^{-1}$, $(6.9\pm2.1)\times10^{36}\ erg\ s^{-1}$ and $(1.7\pm0.9)\times10^{36}\ erg\ s^{-1}$, respectively for all, red and blue GCs. 
If LMXBs in red and blue GCs have the same average intrinsic luminosity, we derive a red/blu ratio $\simeq 3$ of GCs hosting LMXBs ($2.5\pm1.0$ or $4.1\pm2.5$ depending on the surveyed region); alternatively, assuming the fractions observed for brighter sources, we measure an average X-ray luminosity of $L_{X}=(4.3\pm1.3)\times10^{37}\ erg\ s^{-1}$ and $L_{X}=(3.4\pm1.7)\times10^{37}\ erg\ s^{-1}$ per red and blue GC-LMXBs respectively.
In the assumption that the XLF follows a power-law distribution, we find that a low-luminosity break is required at $L_{X}\leq 8\times10^{37}$ erg s$^{-1}$ both in the whole, as well as in the color-selected (red and blue) subsamples. Given the bright-end slopes measured above the X-ray completeness limit, this result is significant at $>3\sigma$ level. Our best estimates for the faint end slope are $\beta_{L}=-1.39/-1.38/-1.36$ for all/red/blue GC-LMXBs. We also find evidence that the luminosity function becomes steeper at luminosities $L_X\gtrsim 3\times 10^{39}$ erg s$^{-1}$, as observed in old ellipticals.}{We conclude that, if most GCs host a single X-ray binary, in NGC 1399 the XLF flattens at low luminosities as observed in other nearer galaxies, and discuss some consequences on LMXBs formation scenarios.}

\maketitle
\section{Introduction}
\label{intro}

X-ray binaries (XRBs) are stellar binary systems consisting of a collapsed object (a neutron star or a black hole) accreting material from a donor star. Beyond being a probe of the physics of the accretion processes, X-ray observations allow us to investigate the nature and evolution of stellar remnants and their host stellar systems. In fact XRBs contribute a significant fraction of the X-ray luminosity of galaxies, and the properties of the XRB population have been shown to be linked to the host galaxy type, its star formation history, galaxy environment and merging history \citep[see][]{Fabbiano89, Fabbiano06}. Several correlations have been found between the total X-ray luminosity from high-mass XRB (HMXB) and the galaxy star formation rate \citep{Ranalli2003, gilfanov2004, Hornschemeier2005, Lehmer2010, Mineo2012} as well as  between the total emission from low-mass XRBs (LMXBs) and the galaxy stellar mass \citep{gilfanov2004, Lehmer2010, Boroson2011, Zhang2012}. Thus XRBs can be used as a probe of the 
assembling process of galaxies over cosmic time \citep{Fragos2013}.  
Furthermore, the study of LMXBs probes the dynamics of dense stellar system such as globular clusters since they form most likely via favorable multi-body encounters. For such reason, the understanding of their formation and evolutionary pathways has strong implications also for different types of studies such as the determination of the number of close binaries, which can end up in a merging event and produce electromagnetic transients as well as gravitational waves.

Given the old stellar population of elliptical galaxies their XRB population is represented by low-mass X-ray binaries. One of the primary tools we use to characterize the XRB population is the X-ray Luminosity Function (XLF), which is dominated by XRBs in the luminosity range detectable in most external galaxies ($>10^{37}$ erg s$^{-1}$). The LMXBs Luminosity Function is steeper than that of young stellar systems \cite[dominated by high-mass X-ray binaries and supernova remnants, see][and references therein]{Fabbiano06}, and has two breaks: a high-luminosity break at $L_{X}\geq 2\times10^{38}$ erg s$^{-1}$ \citep{gilfanov2004, Kim04, Humph08}, which may be absent in younger elliptical galaxies \citep{kim_fabbiano2010},  and a low-luminosity break at $L_{X}\sim5\times10^{37}$ erg s$^{-1}$ \citep{VossGilf06, Kim09, Voss09}. 
The origin of such breaks is still debated, as they can be produced by differences in orbital period, mass ratio, type of donor/accreting star, or  evolutionary stage of the XRBs population. Furthermore, in some elliptical galaxies such features are detected with low statistical significance \citep[see][]{Kim06b} and incompleteness effects may be partly responsible for early reports about the low-luminosity break.

The discovery that in elliptical galaxies a large fraction of LMXBs ($20\%-70\%$) resides in Globular Clusters \citep[GC; ][]{Angelini01, Sarazin03, Kim06a, P11}, has triggered a debate about whether the GC and field populations of LMXBs share the same origin and properties. 
In these galaxies, these binary systems are more likely to be hosted in brighter, more compact and red (high metallicity) GCs, rather than fainter, less compact and blue (metal-poor) GCs \citep{Bellazzini95, Angelini01, Kundu02, Sarazin03, Jordan04, Kim06a, Kundu07, Sivakoff07,P11, Kim13}.
The subpopulation of LMXBs in the field (field-LMXBs) and in Globular Clusters (GC-LMXBs) differs in spatial distribution, as the field-LMXBs follow the parent galaxy stellar distribution, while the GC-LMXBs follow the more extended GC distribution \citep[e.g.][]{Kundu07, P11}; they also differ in the dependence on GC specific frequency $S_{N}$, since the number of LMXBs in GCs depends more strongly on $S_{N}$ than those in the field, as expected if not all binaries are originally formed within GCs \citep{Kim09,P11,Mineo13}. Furthermore, at low luminosities the X-ray luminosity function of GC-LMXBs shows a more pronounced flattening than that of field-LMXBs, due to a lack of faint sources in GCs compared to the field \citep{Kim09,Voss09,Zhang11}.
  
Given that long exposures are needed to study the low-luminosity XLF, these studies have been possible only for a handful of very nearby galaxies. This work is aimed at investigating the faint-end of the XLF of GC-LMXBs in NGC 1399: the cD galaxy at the center of the Fornax cluster, which lies at a distance $D=20.1\pm0.4$ Mpc \citep{dunn06}. The GC-LMXB connection in NGC 1399 was studied in detail by \citet[][hereafter P11]{P11}  using combined \textit{Hubble Space Telescope} (HST) and \textit{Chandra} observations. This galaxy represents an ideal target for studying the GC-LMXBs XLF, since it has the highest fraction of GC-LMXBs among all known nearby galaxies, it is near enough to resolve GC sizes with the \textit{Advanced Camera for Surveys} (ACS), and distant enough to sample efficiently the GC distribution out to large galactocentric radii. 

In P11 the XLF of NGC 1399 was studied in the luminosity range from $\sim8\times10^{37}$ erg s$^{-1}$ up to $\sim10^{39}$ erg s$^{-1}$, where the XLF of both field- and GC-LMXBs is well approximated by a simple power-law, with marginal differences between the two populations (steeper XLF and lower median luminosity for field sources compared to GC ones). Below the lower luminosity limit, incompleteness effects complicate the detection of individual LMXBs.
Here we use the same data as in P11 to perform a stacking experiment in order to probe the presence of faint GC-LMXBs, and to constrain the XLF slope at luminosities below the $8\times 10^{37}$ erg s$^{-1}$ and therefore study the dependence of the faint-end XLF on the color of the host GC.

\section{The dataset}
Our analysis is based on a combination of HST and \textit{Chandra} observations. An extended discussion of the data properties and sample selection can be found  in P11 and Puzia et al. (2014); here we briefly summarize the main properties of the dataset, referring the reader to the latter references for more details.

\subsection{Optical data}
\label{opt_data:sec}
The optical data are based on observations with the HST Advanced Camera for Surveys (ACS; GO-10129, PI: T. Puzia) in the F606W filter. The observations were arranged in a $3\times3$ mosaic centered on the coordinates RA (J2000)$=03^{h}38^{m}28.62^{s}$ and DEC (J2000)$=-35\degree28'18.9''$ (see Fig. 1 in P11), so that the full mosaic extends out to a galactocentric radius of $\sim50$ kpc from the galaxy center, ~$\sim\!5.2$ effective radii of the diffuse galaxy light \citep{RC3} and $\sim\!4.9$ core radii of the globular cluster system density profile \citep{Schuberth10}. 
Each field was observed for a total integration time of 2108 seconds, and the individual exposures were combined into single images using the MultiDrizzle routine \citep{koekemoer02}, with a final pixel scale of $0.03''$/pixel which, at the distance of NCG 1399, corresponds to 2.93 pc ($1'' = 97.7 pc$).  The optical source catalog was created with the Source Extractor software (SExtractor, \cite{Bertin96}), and the astrometric solution was registered to the USNO-B1
catalog\footnote{http://tdc-www.harvard.edu/software/catalogs/ub1.html}, obtaining a final accuracy of 0.2\arcsec\ r.m.s. GCs were selected as sources having a SExtractor stellarity index $\geq0.9$ and $m_{V}(F606W)<26$ mag (in the VEGAmag system), in order to exclude extended sources and compact background galaxies. This results in a completeness of $>80\%$ with $<10\%$ contamination; we refer the reader to P11 for detailed description of how these figures were derived, and to \cite{Brescia12} for a discussion on alternative selection techniques.

We include photometry for our GCs, matching our catalog both with the \cite{Bassino06} $C-T1$ ground based GC catalog and the HST/ACS $g-z$ color catalog from \citet{Kundu05}. We adopted the following color and magnitude cuts to separate the different GC populations:  
\begin{itemize}
\item \textit{Ground-based data}: $1.0\leq C-T1<1.65$ for blue/metal-poor GCs and $1.65\leq C-T1<2.2$ for red/metal-rich GCs, and $T1<23$ mag;
\item \textit{HST data}: $1.3\leq \textit{g-z}<1.9$ for blue/metal-poor GCs, $1.9\leq \textit{g-z}<2.5$ for red/metal-rich GCs, with $z<22.5$ mag (in the VEGAmag system).
\end{itemize}

The $T1$ and $z$ magnitude cuts were adopted to ensure a uniform completeness limit across the whole FOV, as discussed in more detail in P11. 

\subsection{X-ray data}
The X-ray data were retrieved from the \textit{Chandra} public archive\footnote{http://cxc.harvard.edu}. We used the observations with obsID $\#319$ and $\#4172$ (Table \ref{table:observations}) for a total exposure time of  $\sim100$ks. The data were reduced with the CIAO software, extracting standard-grade events in the $0.3\!-\!8$ keV energy band and generating the corresponding exposure maps. As discussed in detail in P11, we used  the {\sc wavdetect} algorithm \citep{Freeman02} to assemble the X-ray source catalog, and the ACIS Extract software \cite[AE,][]{Broos10} to derive the photometric parameters.
AE is designed to deal with a combination of different observations, accounting for variations of source position inside the FOV and over the ACIS detector, using library
templates of the ACIS PSF, as well as to improve the positional accuracy of our catalog. 
In fact the final accuracy of the X-ray catalog is 0.33\arcsec\ with a maximum
systematic offset of 0.6\arcsec.
The properties of all 230 X-ray sources detected individually in our observations are presented in detail in P11. In particular matching the optical and X-ray catalogs yielded 
164 X-ray sources with optical counterparts within 1\arcsec , 
out of which 136 are matched with GC candidates.

\begin{table}[]
\caption{Journal of \textit{Chandra} Observations}            
\label{table:observations}     
\centering    
          \scriptsize{           
\begin{tabular}{c c c c c c}       
\hline\hline               
 Obs. & Detector & Date & RA(J2000) & DEC(J2000) & $T_{exp}$  \\    
\hline                       
$\#319$  & ACIS-S & 2000-01-18 & $03^{h}38^{m}29.4^{s}$ & $-35\degree27'00.4''$ & 56 ks  \\
$\#4172$  & ACIS-I & 2003-05-26 & $03^{h}38^{m}25.6^{s}$ & $-35\degree25'42.6''$ & 45 ks  \\ 
\hline                                  
\end{tabular}}
\end{table}

\section{GC sample selection}
\label{sec:sample_selec}
As described in \S \ref{intro}, the purpose of this work is to find clues on the properties of the GC-LMXB population below the threshold of $L_X\sim 8\times 10^{37}$ erg s$^{-1}$. We tackle this problem by exploiting our optical catalog to stack the X-ray data at the positions of X-ray \textit{undetected} GCs.
In order to maximize the efficiency of our stacking process, we have to take into account the main effects that contribute to lower the S/N ratio of faint LMXBs in our observations, i.e. the X-ray emission of the diffuse gaseous halo of NGC 1399 which affects mainly sources near the galaxy center, and the increase in PSF size with distance from the ACIS aimpoint which is dominant at large galactocentric radii. .
We thus limit our analysis to an annulus around the galaxy center with internal radius $r_{i}=60''$ and external radius $r_{e}=240''$ and consider only GCs with colors in the ranges discussed in \S \ref{opt_data:sec}, which additionally minimizes the contamination from background AGNs. To better constrain the XLF below the individual-source detection threshold (\S \ref{XLF_sec}), we also defined a second smaller sample limited to GCs in the range $120''<r<180''$, which correspond to the region with higher detection efficiency (and lower incompleteness at low luminosities) for point like sources due to the minimisation of the combined effects discussed above, as shown in Figure 7 of P11.

\begin{figure}[]
   \centering
   \includegraphics[width=9cm]{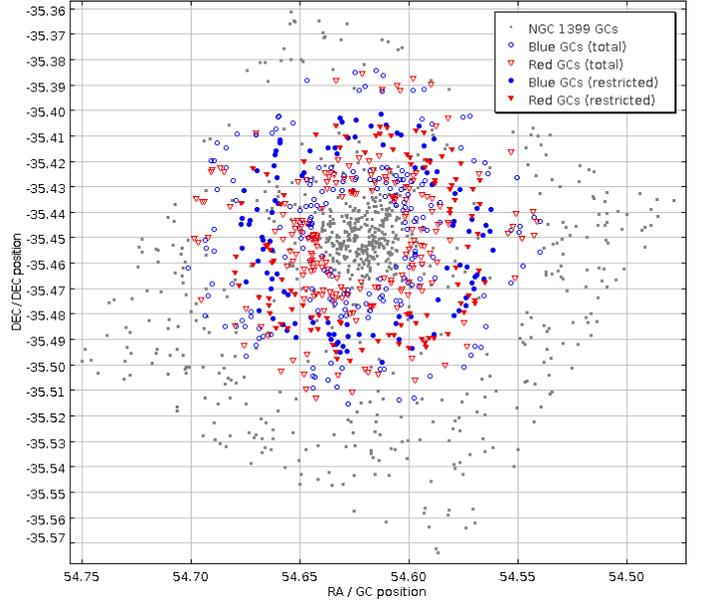}
   \caption{Spatial distribution of the GCs used for the stacking experiments: blue circles and red triangles represent the \textit{total} blue and red samples with no detected X-ray counterpart in P11. Solid symbols identify the \textit{restricted} sample. Gray dots represent the rest of the GC system within the HST/ACS mosaic.}
   \label{spatial_dist}
    \end{figure}

In assembling the input GC position list for X-ray stacking, we excluded from our GC samples, all GCs individually detected in X-rays (Tab. 4 in P11). We further excluded, from the list, all GCs closer than $3\arcsec$ to an individually detected LMXB in order to avoid cross-contamination effects\footnote{within the considered spatial region, the ACIS PSF size is such that more than 90\% of the flux is confined within this radius}. The two catalogs so selected consist of 618 GCs (of which 316 blue and 302 red) in the \textit{total} annulus ($60''<r<240''$) and of 210 GCs (of which 109 blue and 101 red) in the \textit{restricted} annulus ($120''<r<180''$). The X-ray images centered on the input GCs were finally visually inspected to ensure that we did not include any source lurking just below the detection threshold. In Fig. \ref{spatial_dist} we show the spatial distribution of the GC samples used for the stacking experiment.   

\section{X-ray stacking}
As done for individually detected sources, for our stacking experiment we used the AE software to derive the stacked photometry in the 0.5-8 keV energy band for the sample of X-ray undetected GCs described in the previous section.

Given a list of source positions, AE models a polygonal extraction region which approximates the contour of the Chandra-ACIS PSF, taking into account the exact position within the detector; this region is used to measure the aperture photometry of the source and derive the total flux, after applying a  correction for the encircled energy within the extraction area. In addition AE defines, for all nearby objects, a circular mask (with radius 1.1 times a radius that encloses 99\% of the total flux) that is used to exclude the corresponding pixels from the estimation of the source flux and of the local background, in order to prevent the contamination due to close neighbors. This process is repeated for each observation separately, in order to account for the different position and exposure of every source in each image; the final flux is defined as an exposure-weighted average of the single measurements in each observation.
Thus AE uses the appropriate PSF model, deriving the corresponding exposure map for each \textit{Chandra} pointing separately, and generating masks that optimize the source and background extraction regions accounting for nearby objects. This procedure allows to easily adapt AE to stacking analysis if we treat all sources as individual observations of the same object. In such case AE combines the individual photometric measurements, weighting them appropriately by their corresponding exposure time.

\begin{table}
\caption{Average flux and luminosity per GC, measured by the stacking experiment for all sources, as well as the two color subsamples, in the two different annular regions described in the text.}

\label{table:stacking}     
\centering                      
\begin{tabular}{l l c c}       
\hline
sample & no. of & $<f_{X}>_{GC}$ & $<L_{X}>_{GC}$ \\
 & sources & [$10^{-17}$ erg cm$^{-2}$ s$^{-1}$] & [$10^{36}$ erg s$^{-1}$] \\
\hline\hline
\multicolumn{4}{c}{\textit{Total sample:} $60''<r<240''$} \\
\hline
\noalign{\smallskip}
all GCs & 618 &  $13\pm1.6$ & $6.2\pm0.8$ \\
red GCs & 302 & $19\pm2.4$ & $8.9\pm1.0$ \\
blue GCs & 316 & $7.5\pm2.2$ & $3.6\pm1.0$ \\
\noalign{\smallskip}
\hline\hline
\multicolumn{4}{c}{\textit{Restricted sample:} $120''<r<180''$} \\
\hline
\noalign{\smallskip}
all GCs & 210 &  $7.5\pm2.0$ & $3.6\pm1.0$ \\
red GCs & 101 & $14\pm4$ & $6.9\pm2.1$ \\
blue GCs & 109 & $3.6\pm1.9$ & $1.7\pm0.9$ \\
\noalign{\smallskip}
\hline\hline
\end{tabular}
\end{table}

We repeated the stacking experiment six times: in each one of the two annuli discussed in \S \ref{sec:sample_selec}, using the whole GC sample and, separately, for the blue and red populations. The X-ray fluxes and luminosities were derived assuming a bremsstrahlung emission ($kT=7\ keV$) with a photoelectric absorption component $N_{H}=1.3\times10^{20}$ cm$^{-2}$, which mimics the the average spectrum of bright ($L_X\simeq 10^{38}$ erg s$^{-1}$) LMXBs (see P11). The average energy fluxes and luminosities are summarized in Table \ref{table:stacking}. The table shows that we are able to detect an X-ray signal with a significance of $>7.5\sigma$, $9\sigma$ and $> 3.5\sigma$ for the whole, red and blue subsamples respectively in the \textit{total} sample; similarly for the \textit{restricted} sample the significances are $>3.5\sigma$, $>3\sigma$ and $\sim 2\sigma$. Furthermore we find that the average X-ray luminosity of red GCs is larger than that of the blue GCs with more than $3.5\sigma$ and $2\sigma$ 
confidence respectively for the \textit{total} and \textit{restricted} samples.
\begin{figure*}[]
\centering
\includegraphics[width=0.7\textwidth]{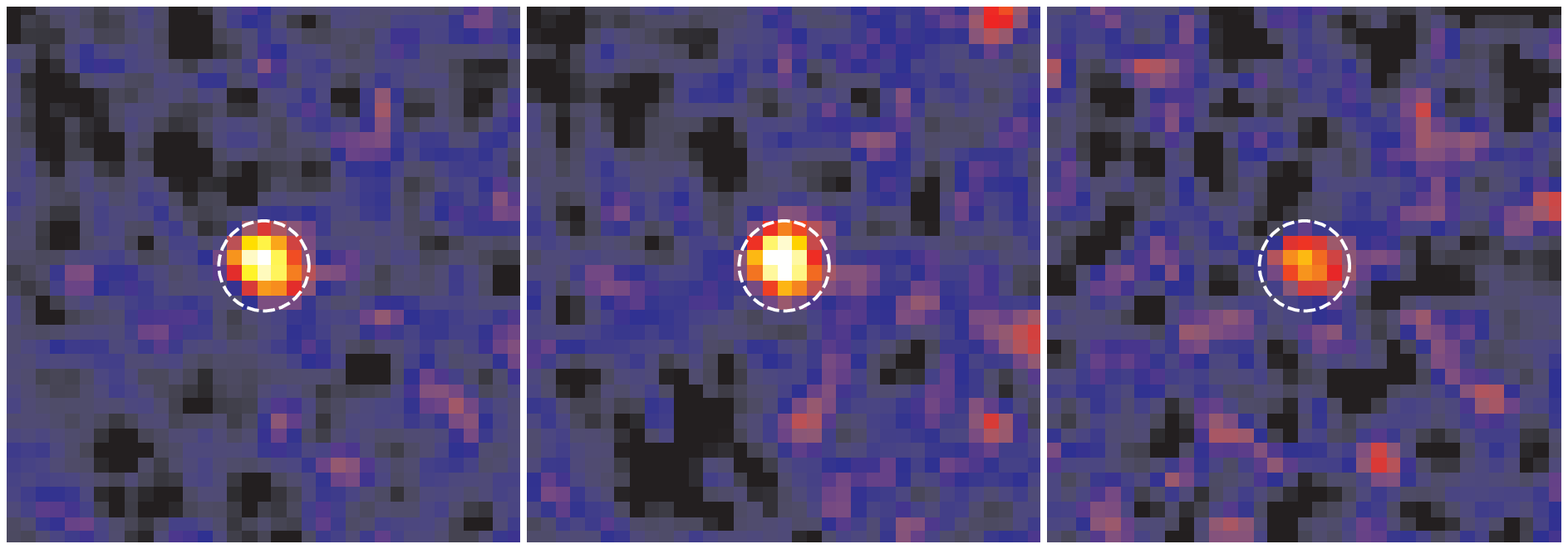}
\includegraphics[width=0.7\textwidth]{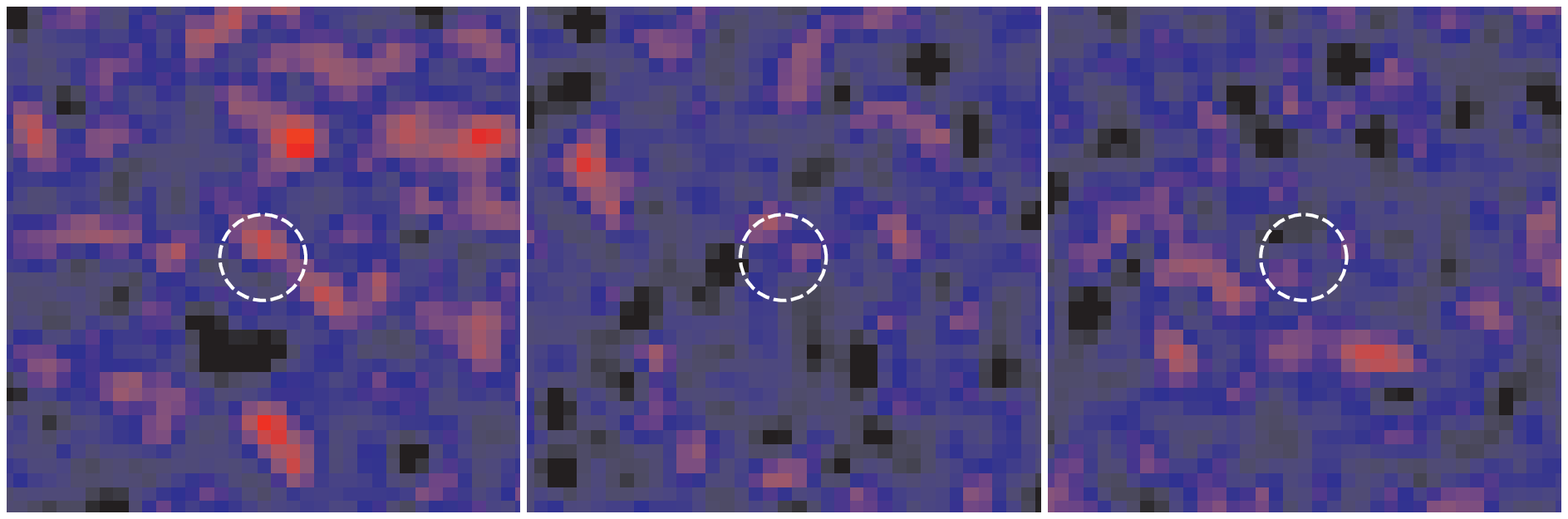}
\caption{\textit{\textbf{Upper row:}} Stacked images of all GCs subsamples: \textit{left}) all 618 GCs; \textit{center}) 316 red GCs; \textit{right}) 302 blue GCs. The central circle highlights the 3\arcsec diameter region where most of the flux from the stacked LMXBs is concentrated. The images are produced through a sigma-clipped averaging algorithm with the clipping threshold set at $4\sigma$. \textit{\textbf{Lower row:}} Same as the upper row, for 3 of the 100 samples simulated using random positions near real GCs (see discussion in text).}
\label{stacked_images}
\end{figure*}

In Fig. \ref{stacked_images} (upper row) we show the stacked images of the neighborhood of each of the three color subsamples for the annulus $60''<r<240''$, obtained computing the average value per pixel after applying a sigma clipping threshold of 4$\sigma$. These images are presented here only to visualize the results of the stacking process, since the actual flux measurements reported in Table \ref{table:stacking} are derived by AE through the more accurate procedure discussed above. As an additional verification of our method, we measured the total flux from the images in Fig. \ref{stacked_images}, finding that the results are fully consistent with those produced by AE. Each input GC position was also visually inspected to make sure that there is no evident X-ray source lurking just below the detection threshold used by P11.

In order to test the robustness of our results, we performed 100 simulations repeating the stacking process at random positions around each GC. Each simulation contained as many simulated positions as the actual dataset; these positions were placed between $3\arcsec$ and $15\arcsec$ from the actual GCs in order to preserve the source and background spatial distributions, preventing any contamination from the GCs themselves. As done for the real stacking experiment, we excluded any position closer than $3\arcsec$ to a detected X-ray source (whose pixels are anyway masked by the AE algorithm) and we forced the simulated positions to be at least $3\arcsec$ apart from each other to make sure that each simulation represents an independent dataset. We find an average stacked luminosity of $L_X=(1.6\pm 2.6)\times 10^{38}$ erg s$^{-1}$, i.e. $<L_{X}>_{GC}=(0.8\pm 1.2)\times 10^{36}$ erg s$^{-1}$ consistent with a null flux, thus confirming for real GCs a significant detection at $99\%$ level. Some examples of the 
result of the stacking experiment for the simulated sample is shown in the lower panel of Fig. \ref{stacked_images}.

Since our stacking process only allows us to derive average luminosities, to compare our result with the properties of the brighter population of sources detected individually in P11, we need to make some assumptions on the distribution of LMXBs within GCs. Assuming that red and blue GC-LMXBs have the same average X-ray luminosity distribution function, the relative number of red ($\phi_{red}^{GC-LMXB}$) and blue ($\phi_{blue}^{GC-LMXB}$) GCs hosting a LMXB, for our \textit{total} sample, is given by:


\begin{equation}
\frac{\phi_{red}^{GC-LMXB}}{\phi_{blue}^{GC-LMXB}}=\frac{L_{red}^{stack}}{n_{red}}\ \frac{n_{blue}}{L_{blue}^{stack}}=\frac{<L_{X}>_{red GC}}{<L_{X}>_{blue GC}}=2.5\pm1.0
\end{equation}

\noindent where $L_{red}^{stack}$ and $L_{blue}^{stack}$ are respectively the stacked luminosities obtained from the red and blue GC subsamples while $n_{red}$ and $n_{blue}$ are the number of the GCs in such subsamples.
A similar calculation for the \textit{restricted} sample yields:
\begin{equation}
\frac{\phi_{red}^{GC-LMXB}}{\phi_{blue}^{GC-LMXB}}=4.1\pm2.5
\end{equation}

In P11 we found that the fractions of red ($\phi_{red}^{GC-LMXB}=16\%$) and blue ($\phi_{blue}^{GC-LMXB}=5\%$) GCs hosting a LMXB lead to a ratio $\phi_{red}^{GC-LMXB}/\phi_{blue}^{GC-LMXB}=3.2$.
The ratio for the stacked sample is thus consistent within $1\sigma$ with the ratios observed at higher luminosities (and over the whole galaxy) in NGC 1399, and in agreement with the 3:1 ratio usually reported in the literature for other elliptical galaxies \citep{Kundu02, Jordan04, Kim06a, Kundu07, Sivakoff07, Kim13}. 

Conversely, fixing the relative fractions $\phi_{red}^{GC-LMXB}$ and $\phi_{blue}^{GC-LMXB}$ to the values observed at brighter luminosities, we find that faint red and blue GC-LMXBs, in the \textit{total} sample, have compatible average luminosities:
\begin{eqnarray}
\overline{L}_{red}=\frac{L_{red}^{stack}}{n_{red}\cdot{\phi_{red}^{GC-LMXB}}}=(5.5\pm 0.7) \times10^{37}~\mbox{erg s}^{-1}\\
 \overline{L}_{blue}=\frac{L_{blue}^{stack}}{n_{blue}\cdot{\phi_{blue}^{GC-LMXB}}}= (7.2\pm 2.0) \times10^{37}~\mbox{erg s}^{-1}
\end{eqnarray}
The same holds for the \textit{restricted} sample, where:\smallskip\\
\smallskip
$\overline{L}_{red}=(4.3\pm 1.3)\times 10^{37}~\mbox{erg s}^{-1}$\\
$\overline{L}_{blue}=(3.4 \pm 1.7)\times 10^{37}~\mbox{erg s}^{-1}$.\\

In both cases, the average  luminosity is consistent with an emission below the completeness limit of the (uncorrected) X-ray data of $L_X\sim2\times 10^{38}~\mbox{erg s}^{-1}$ found in P11. Moreover for the \textit{restricted} sample where, as discussed before, the source catalog reaches fainter completeness levels, the stacked luminosities are lower than the faintest source found in P11 with $L_X\sim 5\times 10^{37}$ erg s$^{-1}$ , as expected if we are detecting the cumulative emission due to LMXBs below our individual source detection threshold.


\section{The GC-LMXBs luminosity function}
\label{XLF_sec}
As discussed in the introduction, previous studies have found that the GC-LMXBs X-ray Luminosity Function in elliptical galaxies can be represented by a double power-law with a break at $\sim5\times10^{37}\ \mbox{erg s}^{-1}$. This result is limited so far only to a few nearby galaxies, for which very long exposures enable the detection of individual LMXBs down to X-ray luminosities of $10^{35}$ erg s$^{-1}$. Our stacking analysis allows us to verify whether this result holds also for a giant elliptical as NGC 1399. To this end we parametrize the XLF as:
\begin{equation}
\label{XLF_eq}
dN/dL=\Phi (L) = \left\{ \begin{array}{rl} 
A \cdot (L/L_{bk})^{\beta_{L}} & \mbox{ if $L<L_{bk}$} \\
A \cdot (L/L_{bk})^{\beta_{H}} & \mbox{ if $L\ge L_{bk}$}
\end{array}  \right.
\end{equation} 
where $A$ is the normalization of the XLF at the break luminosity $L_{bk}$, while $\beta_{L}$ and $\beta_{H}$ are respectively the faint end slope and the bright end slope of the XLF.

\begin{figure*}
 \centering
\includegraphics[width=8cm]{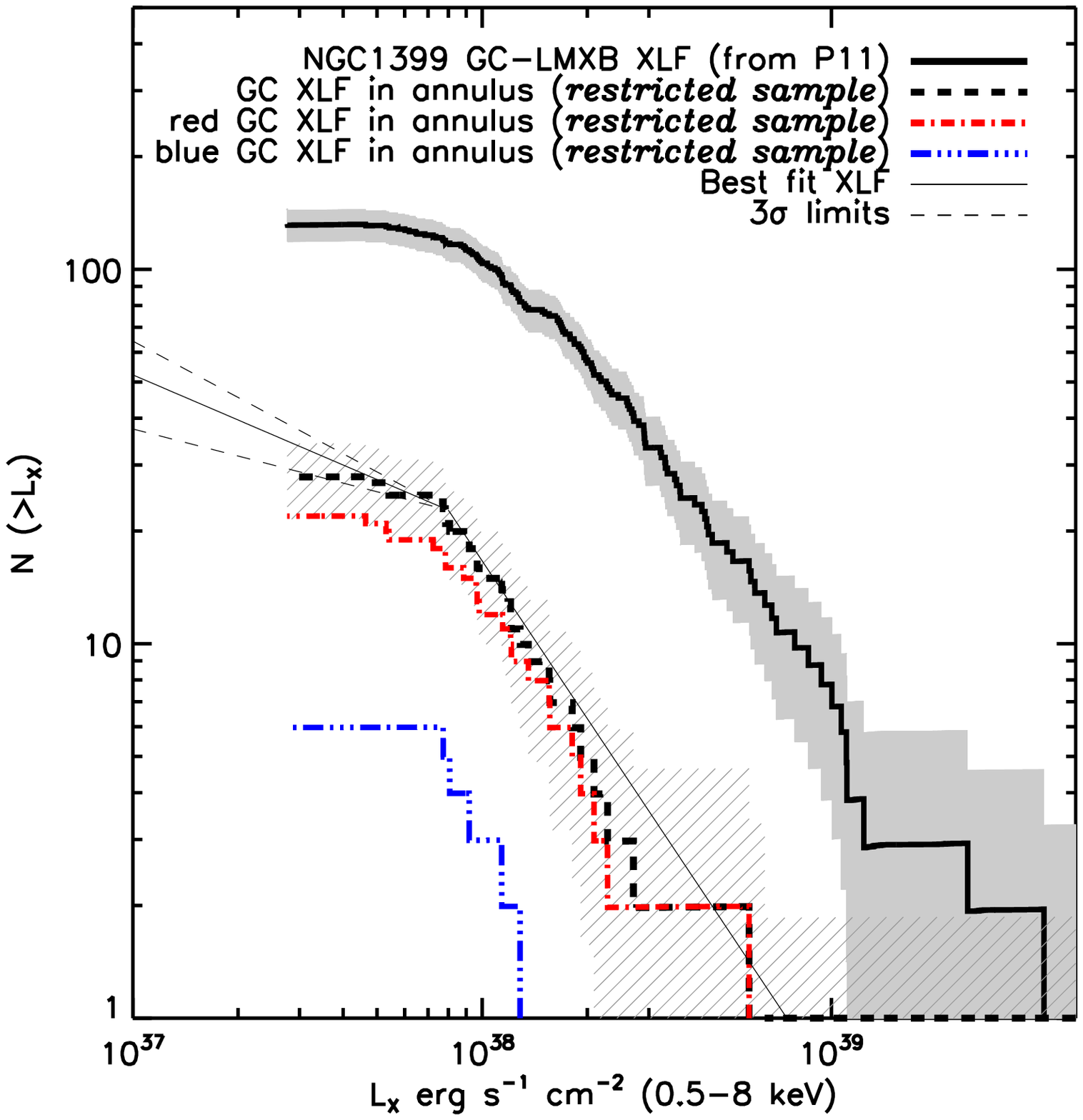}
\includegraphics[width=8cm]{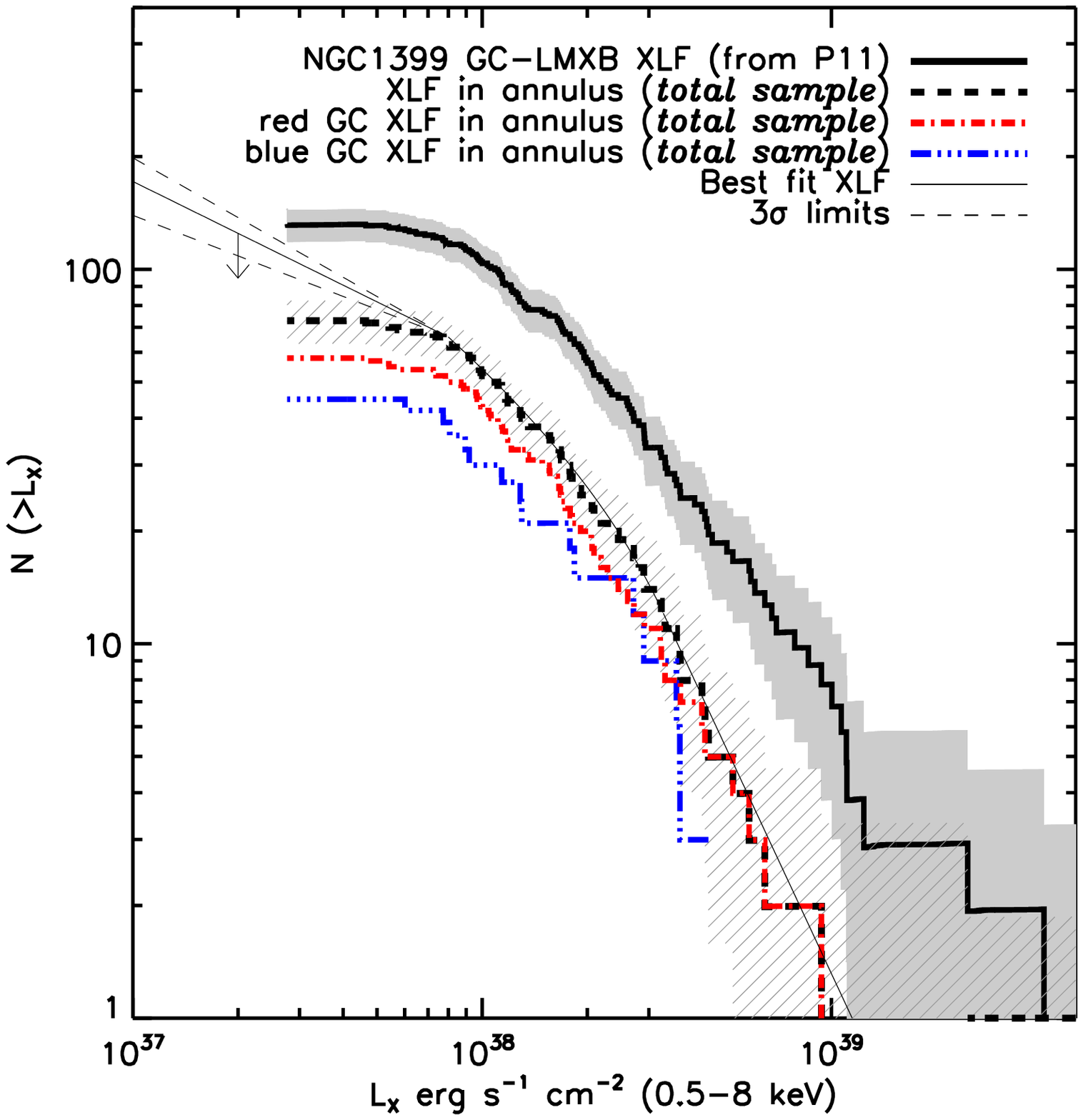}
  
\caption{\textbf{\textit{Left panel:}} Cumulative XLF for all, red and blue (individually detected) GCs-LMXB within the $120\arcsec-180\arcsec$ annulus (\textit{restricted} sample). For comparison the thick solid line represents the total XLF in NGC 1399, as derived in P11. The statistical uncertainties are shown, for the total NGC 1399 and annular XLF, as shaded (solid and dashed) areas. 
 For clarity, the best-fit analytical XLF (according to equation \ref{XLF_eq}), is shown only for the total GCs-LMXB  annular sample as a thin solid line, while its upper and lower $3\sigma$ limits (Table \ref{table:LF}) are marked by dashed lines. \textbf{\textit{Right panel:}} as left panel for the \textit{total} sample. The downward arrow emphasises that in this case the faint-end slope has to be considered a conservative limit (see discussion in the text).}
 \label{all_LF}
\end{figure*}

The XLF of bright, individually detected sources was already presented in P11: here we refine the analysis, considering the two radial annuli used to define the \textit{total} and \textit{restricted} samples in \S \ref{sec:sample_selec}. We point out that for the study of the features of the XLF, the \textit{restricted} sample allows to derive more stringent constraints than the \textit{total} one: in fact as noted before the annulus $120''<r<180''$ represents the region less affected by incompleteness problems, allowing to probe the X-ray population down to fainter luminosities. For such reason we start discussing the \textit{restricted} sample, and later compare the results to those obtained from the wider annulus.
To this end we first fitted the luminosity function of the \textit{individually detected} sources in the $120\arcsec - 180\arcsec$ range, with a power law function; we used the Maximum Likelihood (ML) plus Kolmogorov-Smirnov (KS) approach described by \cite{Clauset09, Newman2005}, in order to derive both the power law slope, its uncertainty and the minimum luminosity above which the data follow a power law distribution. We obtain a differential XLF best-fit slope of  $\beta_H=-2.4\pm 0.3$ above a minimum luminosity of $L_X=8\times 10^{37}$ erg/s (Figure \ref{all_LF}). 

We note that this slope is steeper than the one found for the total GC XLF in P11, which had $\beta_H=-1.7\pm 0.2$. However this discrepancy (anyway consistent within the errors) can be explained as a combination of different factors: first of all the total XLF suffers from completeness issues at the faint end due to the problems discussed in $\S \ref{sec:sample_selec}$ which tend to produce a flatter distribution; while a completeness correction was implemented in P11, residual effects are likely still present. Second, in the case of the total XLF the fitting method was a simple $\chi^2$ due to the need to correct for AGN contamination (an effect which is not significant here because of the smaller area), and this generally results in lower accuracy fits. Some residual AGN contamination was also possibly present at the bright end since the total sample was not color-selected as here, (see discussion in P11), again resulting in a flatter XLF slope, and finally we cannot exclude that the XLF is intrinsically 
different at different galactocentric distances. We thus adopt here the $\beta_H=-2.4$ value obtained from the ML fit. 

We did not detect a break in P11, although the GC XLF seems to flatten below $\sim 8\times10^{37}\ \mbox{erg s}^{-1}$; in fact, as discussed above, completeness effects may be partly responsible for the flattening. A similar bend is observed here for the $120\arcsec - 180\arcsec$ XLF.  We thus conservatively assume that $L_{bk}=8\times10^{37} \mbox{erg s}^{-1}$, and we will discuss below the consequences of relaxing such hypothesis. Note that this break value is also supported by the KS test performed to identify in what luminosity range a power law function provides the best fit.

The faint-end slope $\beta_{L}$ is determined requiring that the integral of the XLF below the break is equal to the total measured luminosity for $L<L_{bk}$. The latter is the sum of two components: the 
stacked luminosity $L_{stack}=<L_{X}>_{GC}\cdot n_{GC}$ presented in Table \ref{table:stacking}, and the residual contribution $L_{faint LMXB}$ of the 7 individually detected LMXBs with $5\times 10^{37}$ erg s$^{-1}<L<L_{bk}$:
\begin{equation}
\label{eq:XLF_int}
\int_{10^{32}}^{8\times10^{37}}\Phi(L)LdL=L_{stack}+L_{faint LMXB}
\end{equation}
 where for the lower integration limit we adopted the average luminosity of a quiescent binary star $L=10^{32}$ erg s$^{-1}$(Verbunt \& Lewin, 2004); however, even if we integrate the function down to $L_X=0$, our results do not change significantly. The results, for all the subsamples discussed in the previous section, are summarized in Table \ref{table:LF} while in Fig. \ref{all_LF} is shown the differential GC XLF $\Phi(L)$ and its $3\sigma$ limits. 
The effect of the 12 GCs that we removed due to the overlap with nearby LMXBs (\S \ref{sec:sample_selec}) would produce a slight bias in the determination of the faint end slope but we verified that, assuming that they have the same properties and average X-ray luminosity of the rest of the stacked GC sample, the faint-end slope would increase by only $\sim 1\%$.

\begin{table}[h]
\caption{Differential slope $\beta_L$ of the faint-end of the GC XLF for the \textit{total} and \textit{restricted} samples, derived from equation \ref{eq:XLF_int}, with upper and lower $3\sigma$ limits}             
\label{table:LF}      
\centering                       
\begin{tabular}{c c c c}        
\hline\hline                 
  & $\beta_{L}$ & $\beta_{L}+3\sigma$ & $\beta_{L}-3\sigma$ \\    
\hline\hline                        
\multicolumn{4}{c}{\textit{Total sample:} $60''<r<240''$} \\
\hline
All GCs  & $-1.46$ & $-1.36$ & $-1.53$  \\
Red GCs  & $-1.44$ & $-1.36$ & $-1.50$  \\
Blue GCs & $-1.53$ & $-1.23$ & $-1.67$  \\
\hline\hline                                   
\multicolumn{4}{c}{\textit{Restricted sample:} $120''<r<180''$} \\
\hline
All GCs  & $-1.39$ & $-1.23$ & $-1.49$  \\
Red GCs  & $-1.38$ & $-1.21$ & $-1.48$  \\
Blue GCs & $-1.36$ & $indef.$\tablefootmark{a} & $-1.54$  \\
\hline\hline
\end{tabular}
\tablefoot{\tablefoottext{a}{The lower limit on the differential XLF slope of blue GCs in the \textit{restricted} sample is undefined since the stacking results are compatible within $3\sigma$ with a null flux and a flat cumulative XLF.}}
\end{table}

The results show that the XLF flattens significantly for $L<L_{bk}$, and that even at the $3\sigma$ confidence level the GC XLF slope is incompatible with the bright end slope ($\beta_H=-2.4$). This conclusion still holds even considering separately the red and blue subsamples; this is not surprising for red GCs which represent the bulk of the population, while for blue GCs the result is confirmed despite the fact that the uncertainties are larger due to the poor statistics. In this latter case the slope of the bright end is very poorly constrained, so we assume that the XLF of blue GCs does not differ from the total one, a conclusion which was supported by the analysis presented in P11 for the whole LMXBs population. 

In deriving the values presented in Table \ref{table:LF} we assumed a break luminosity of $L_{bk}=8\times10^{37} \mbox{erg s}^{-1}$, which is the point where the XLF of \textit{individually detected} sources starts to deviate from a simple power-law (see the solid and dot dashed lines in Figure \ref{all_LF}). It is possible that $L_{bk}$ is actually fainter than this value, and that the flattening of the faint end of the XLF in Figure \ref{all_LF} is due to the completeness limit of our observations (since we are missing an increasing fraction of sources at low fluxes). In such case however our conclusion about the flattening of the faint end slope of the \textit{intinsic} XLF (as opposed to the observed one based on individual detections) at very low luminosities, would be strengthened, since this would imply that a larger fraction of the X-ray flux detected in our stacking experiment is produced in relatively bright LMXBs just below the individual detection threshold. As a result the faint 
end of the XLF would need to be even shallower, if not with a (differential) positive slope, to compensate for the additional flux near the completeness limit.

We note that a deviation from a simple power law seems to be present also at the bright end ($L_X\gtrsim 2\times 10^{38}$ erg s$^{-1}$), where the XLF appears slightly steeper than the model. Given the low statistics it is difficult to tell whether this feature is real, but the same feature is present in the \textit{total} sample as well (as discussed below in more detail); if confirmed, this would agree with the finding by \citet{kim_fabbiano2010} that old ellipticals tend to have a lower fraction of bright LMXBs, compared to younger galaxies, and thus steeper XLFs above a few $\times 10^{38}$  erg s$^{-1}$.

More care is needed to interpret the results derived for the \textit{total} sample since, although in principle we could expect more robust constrains due to the larger statistics, the $60\arcsec - 240\arcsec$ region is severely affected by incompleteness issues. We first note that the XLF fails to follow a single power-law down to $L_X\sim 8\times 10^{37}~\mbox{erg s}^{-1}$ as in the case of the \textit{restricted} sample: the same ML fitting approach used above in this case suggests that the XLF already bends at $L_X\sim 3\times 10^{38}~\mbox{erg s}^{-1}$. Since at this luminosity the XLF is unlikely to be incomplete, as shown in P11 (see their Figure 6 where the corrected and uncorrected XLF are shown), the data seem to confirm the presence of a high-luminosity break as observed in the \textit{restricted} sample. We thus require a more complex model to fit the XLF of \textit{individually detected} sources: we adopt a double power-law with a high luminosity break (as opposed to the low-luminosity 
one discussed earlier) as done in other works \citep[e.g.][]{Sarazin01,Kim09} obtaining a (very high) slope of $\beta_{VH}=-3\pm 0.5$ above the high-luminosity break and $\beta_H=-1.6\pm 0.4$ below it down to $L_{bk}=8\times 10^{37}~\mbox{erg s}^{-1}$. We further note that the position of a low-luminosity break is now more poorly constrained since incompleteness effects become increasingly more important at faint luminosities and part of the XLF flattening is due to incompleteness. Fixing the low-luminosity break to $L_{bk}=8\times 10^{37}~\mbox{erg s}^{-1}$, the stacked flux reported in Table \ref{table:stacking} (plus the contribution of the detected sources below the break) implies that the slope of the faint end is $\beta_L=-1.46$ with the $3\sigma$ limits presented in Table \ref{table:LF}. This confirms the flattening of the XLF at faint luminosities already found using the \textit{restricted} sample. Since the slope of the XLF in the range $8\times 10^{37}~\mbox{erg s}^{-1}<L_X<3\times 10^{38}~\mbox{
erg s}^{-1}$ represents a lower limit to the actual XLF slope, this result is a conservative estimate; in fact the statistical errors quoted here do not account for the systematic uncertainties due to the unknown luminosity of the break nor for the fact that some of the stacked flux may come from sources with luminosity above the $8\times 10^{37}~\mbox{erg s}^{-1}$ limit. These considerations make the results for the \textit{restricted} sample more stringent than the ones for the \textit{total} set.

\section{DISCUSSION AND CONCLUSIONS}
\label{discussion:sec}


The results presented in the previous sections shows that NGC 1399 possesses a population of LMXBs with luminosities below the detection threshold   of individual sources. To be more precise we are able to verify the presence of such population inside GCs (which in fact host the majority of LMXBs in this galaxy, see P11), but it is very likely that follow up observations, reaching fainter detection limits will be able to detect such population also in the field.
This result is in agreement with the direct detection of faint LMXBs by \citet{Kim06b,Fabbiano07,Kim09,Zhang11} in several other nearer ellipticals (Maffei I, Centaurus A,  NGC 3379, NGC 4697, NGC 4278) as well as by \citet{VossGilf07,Revnistev08,Zhang11} for the Milky Way, M31 and M81. On the other hand \citet{Vulic12}, using a stacking method similar to our own, found no evidence of a faint XRB population in M51 stellar clusters, even considering old clusters similar to GCs; however it is not clear how these stellar clusters directly compare with the compact GCs found in ellipticals which are more likely to produce close binaries due to dynamical effects and high central densities (for instance Vulic and collaborators do not detect a dependence on GC luminosity as the one found in \citealt{Sivakoff07} or P11). Furthermore it is difficult to estimate the luminosity range covered in their stacking analysis, since the completeness limit of the detected sources that they exclude from the 
stacking process is not explicitly reported; thus if their stacking includes only sources at very low luminosities ($<10^{35}~\mbox{ erg s}^{-1}$ as stated in the abstract) their data could be consistent with ours if our stacked signal comes from XRBs in the range $10^{35} - 10^{37}$ erg s$^{-1}$. The optical catalog used by Vulic and collaborators is also deeper than ours (the NGC 1399 color-selected GC sample probes $M_V<-7.5$, see P11) which may result in a dilution of the stacked sample if LMXBs are hosted primarily by bright clusters (as generally observed for GCs).

Our analysis shows that the faint GC-LMXBs population has similar properties to its bright counterpart. In particular  we observe that red GCs have an average X-ray luminosity $\sim 3$ times larger than blue GCs ($2.5\pm1.0$ for the \textit{total} sample and $4.1\pm2.5$ for the \textit{restricted} one). Assuming a similar intrinsic average luminosity distribution for all LMXBs, we can interpret this result concluding that red GCs are 3 times more likely to host faint LMXBs than blue ones, as already observed at brighter X-ray luminosities in many recent works (see references in \S \ref{intro}); alternatively, assuming the same likelihood ratio, faint LMXBs have similar average luminosities in red and blue GCs. These findings thus suggest that the same dynamical and chemical effects, which affect the formation and evolution of bright LMXBs, influence the properties of the entire GC-LMXBs population. It is intriguing that a similar abundance ratio/average luminosity holds at every luminosity since at 
high and low $L_X$ the 
formation of close binaries follows different paths and the donor 
star is likely 
to be different \citep{Fragos09,Ivanova13}. In fact for $L_X< 2\times 10^{37}$ magnetic braking in main-sequence stars is the most likely cause of the overabundance of LMXBs in red GCs \citep{Ivanova06}, while at brighter luminosities the metallicity influences the evolution of red-giant donors \citep{Ivanova12}. 
Since our stacked samples can include in principle sources up to $L_X < 8\times 10^{37}$ erg s$^{-1}$ (our individual source completeness limit in the \textit{restricted} sample), we are probing both regimes above and below the transition luminosity, which may explain why we retrieve a similar ratio as for brighter LMXBs, if our stacked luminosity is dominated by sources just below the individual detection threshold. However we point out that our result is consistent with \cite{Kim13} who find that even for $10^{36}<L_X < 2\times 10^{37}$ erg s$^{-1}$ the red vs blue ratio is in the range $1.2-3.7$ in a stacked sample of elliptical galaxies in Virgo and Fornax clusters. One more alternative is that LMXBs in red GCs are simply 3 times more luminous on average than their blue GC counterparts; we note however that such effect is not observed at brighter luminosities where LMXBs are individually detected and their luminosities and spectrum appear consistent within the two subpopulations \cite[see][]{Kim06a,P11}. 
 For the reasons discussed above, this scenario appears less likely as it would imply a difference in the type of faint LMXBs hosted by red and blue GCs, which should also mimic the difference observed in their brighter counterparts.

Our stacking experiment indicates that the XLF of LMXBs flattens at low luminosities in NGC 1399, as already found in other nearer galaxies \citep{VossGilf06, Fabbiano07, Kim09, Voss09, Zhang11} through both direct X-ray binary detection and stacking analysis. On the nature of this break there are various interpretations: for instance the change of the binary braking mechanism from magnetic stellar winds at high luminosity to gravitational radiation at low luminosity \citep{postnov_kuranov2005}, the transition from LMXBs with a predominantly giant donors to binaries with predominantly main-sequence donor \citep{Revnistev11}, or the presence  at low luminosities of a bump, related to contribution of the Red Giant LMXBs population \citep{Kim09}. 
If, as suggested by such works, the XLF of field LMXBs has instead a different shape with steeper slope, this result would argue in favor of a separate formation and/or evolutionary path for GC and field sources \citep[see, e.g, the discussion in][]{Kim09}. 

The hypothesis that LMXBs in GCs and in the field are formed independently (at least most of them) and may represent separate populations is also supported by several studies on spatial profile and specific frequency $S_N$ \citep[e.g.][]{Irwin06,Kundu07,P11}. In this respect we point out that in \cite{P11}, we discussed the link between LMXB abundance in GCs and GC specific frequency (see their Figure 18); in that case we had to extrapolate the NGC 1399 LF down to $L_X=10^{37}$ erg s$^{-1}$ to compare with the results of \cite{Kim09} for NGC 3379, NGC 4697, and NGC 4278. The results shown here indicate that in such comparison we have to use the lower limits for NGC 1399, corresponding to a broken power-law XLF, thus reducing the disagreement between NGC 1399 and other galaxies of similar $S_N$. This indicates that the large number of LMXBs in NGC 1399 is consistent with simple statistical fluctuations in the average elliptical galaxy population, a result confirmed by the comparison made by \cite{Mineo13} at 
brighter luminosities (see their Figure 10). 
On the other hand the same argument would further reduce the dependence of the abundance of field LMXBs on $S_N$, thus arguing against a GC origin of field LMXBs. 

An alternative explanation of the shape of the GC XLF attributes the flattening at low luminosity just to a bias induced by the presence of multiple LMXBs in the most massive GCs; such effect would remove sources from the faint-end of the XLF and increase the number of bright counterparts. While it is plausible that some of the X-ray bright GCs contain multiple binaries (in fact we find evidence for at least one GC in NGC 1399 in P11), variability measurements suggest that a significant fraction of GCs are dominated by a single XRB both in NGC 1399 (P11) and in other ellipticals \cite[see for instance the discussion in][]{Kim09}.

Finally we found evidence that the XLF may not be well represented by a single power-law model over the entire luminosity range above the low-luminosity break, and that a second break is required at $L_X\sim 3\times 10^{38}$ erg s$^{-1}$. This result agrees with the model proposed in earlier works by, e.g., \citet{Sarazin01,Kim09} and may confirm that old ellipticals tend to have a lower fraction of bright LMXBs compared to younger galaxies \citep{kim_fabbiano2010}. This effect was not observed in P11 most likely due to the contamination of background sources (AGNs) which becomes more relevant in the galaxy outskirts. In the present study the contamination is reduced by the smaller surveyed area and the use of color-selected GC subsamples: we estimate, based on extrapolation from deep X-ray counts (see discussion in P11), $<1$ contaminant source over the small annuli considered here. In any case such contaminants would again strengthen our result about the flattening of the XLF at faint luminosities, as do 
all other possible systematics discussed in the previous sections.

Hopefully, deeper \textit{Chandra} observations will allow us in the future to directly probe the faint LMXBs population, and confirm the results discussed above. 
 
\begin{acknowledgements}
M.P. acknowledges support from FARO 2011 project from the University Federico II of Naples. He also thanks the International Academic Exchange Fund of the "Vicerrector\'{i}a Acad\'{e}mica" at the Pontificia Universidad Catolica in Santiago, and the Department of Astronomy and Astrophysics providing travel and lodging support. G.F. is grateful for the hospitality of the Aspen Center for Physics supported by NSF grant no.1066293i.
\end{acknowledgements}


\end{document}